\documentclass[12pt]{article}
\usepackage{epsf,amsfonts,latexsym}  
\newsymbol\ltimes 226E
\newsymbol\rtimes 226F
\epsfverbosetrue
\def\double{\mathbb}         
\def\ccal{\cal}           

\def\cc{{\double C}}     
       
\def\rr{{\double R}}     
\def\zz{{\double Z}}

\def\aa{{\cal A}}
\def\bbb{{\cal B}}
\def\ccc{{\cal C}}
\def\dd{{\cal D}} 
        
\def\hh{{\cal H}}
\def\hhh{{{\double H}}}   

\def\mm{{{\ccal M}}}

\def\aa{{\cal A}}
\def\dd{{\cal D}} 
\def\hh{{\cal H}}

\def\lll{{\cal L}}
\def\sss{{\cal S}}
\def\jj{{\cal J}}
\def\t{{\rm tr}\,}

\def\ddd{{\,\hbox{$\partial\!\!\!/$}}}

\def\de{{\rm d}} 

\def\pa{{\partial}}

\def\lb{\left[} 
\def\rb{\right]}

\def\ot{\otimes}
\def\op{\oplus}

\def\bb{\begin{eqnarray}}
\def\ee{\end{eqnarray}}
\def\eee{\nonumber\end{eqnarray}}
\def\pp{\pmatrix}
\def\qq{\quad}

\begin{document}

\hsize 17truecm
\vsize 24truecm
\font\twelve=cmbx10 at 13pt
\font\eightrm=cmr8
\baselineskip 18pt

\begin{titlepage}

\centerline{\twelve CENTRE DE PHYSIQUE TH\'EORIQUE}
\centerline{\twelve CNRS - Luminy, Case 907}
\centerline{\twelve 13288 Marseille Cedex 9}
\vskip 3truecm

\centerline{\twelve SPIN GROUP AND ALMOST
COMMUTATIVE GEOMETRY}

\bigskip

\begin{center} 
\bf Thomas SCH\"UCKER
\footnote{\, and Universit\'e de Provence,
\qq {\tt schucker@cpt.univ-mrs.fr}} \\

\end{center}

\vskip 2truecm
\leftskip=1cm
\rightskip=1cm
\centerline{\bf Abstract} 

\medskip

For Connes' spectral triples, the group of
automorphisms lifted to the Hilbert space is defined
and used to fluctuate the metric. A few commutative
examples are presented including Chamseddine and
Connes' spectral unification of gravity and
electromagnetism. One almost commutative example
is treated: the full standard model. Here the lifted
automorphisms explain O'Raifeartaigh's
reduction  $SU(2)\times U(3)/\zz_2.$

\vskip 1truecm PACS-92: 11.15 Gauge field theories\\ 
\indent MSC-91: 81T13 Yang-Mills and other gauge
theories 
 
\vskip 1truecm

\noindent july 2000
\vskip 1truecm
\noindent CPT-00/P.4031\\
\noindent hep-th/yymmxxx
 
\vskip1truecm

 \end{titlepage}

\section{Introduction}

The standard model of electro-weak and strong forces
remains the most painful humiliation of theoretical
physics. It is hard to believe that its theoretical
{\bf input},
\begin{itemize}\item
the compact, real Lie group $G=SU(2)\times U(1)\times
SU(3)/(\zz_2\times\zz_3)
$ for the gauge bosons,
\item
the three unitary representations $\hh_L,\ \hh_R,\
\hh_S$ for left- and right-handed spinors and the
Higgs scalars,
\bb\hh_L &=& \bigoplus_1^3\lb
(2,{\textstyle\frac{1}{6}},3)\op
(2,-{\textstyle\frac{1}{2}},1)
\rb  \label{hl},\\  
 \hh_R& = &\bigoplus_1^3\lb 
(1,{\textstyle\frac{2}{3}},3)\oplus
(1,-{\textstyle\frac{1}{3}},3)\op (1,-1,1)
\rb, \label{hr}
\\     
 \hh_S &= &(2,-{\textstyle\frac{1}{2}},1)
\label{hs},\ee   
where $(n_2,y,n_3)$
denotes the tensor product of an $n_2$ dimensional
representation of $SU(2)$, the one dimensional
representation of $U(1)$ with hypercharge $y$:  
$\rho(\exp (i\theta)) = \exp (iy\theta) $ and an $n_3$ 
dimensional
representation of $SU(3)$. For
historical reasons the hypercharge is an integer
multiple of ${\textstyle\frac{1}{6}}$.
\item
 18 real constants: 3 gauge couplings $g_2,g_1,g_3$, 2
scalar couplings $\lambda, \mu$, 13 Yukawa
couplings parameterizing the fermions mass matrix, 
\end{itemize}
and its
complicated {\bf rules} deriving from five action
terms:
\begin{itemize}\item
 the Yang-Mills action, 
\item
the Dirac action,
\item
the Klein-Gordon action,
\item
the Higgs potential,
\item
the Yukawa terms,
\end{itemize}
encode a fundamental theory. Since several
decades experiments continue to confirm the standard
model with ever increasing accuracy. Simultaneously
all theoretical attempts to lessen the humiliation,
technicolour, left-right symmetric models, grand
unification, supersymmetry, supergravity,
superstrings,... have only added to the humiliation, all
attempts except one: Connes'. Like
Minkowskian geometry induces the magnetic force
from the electric force, Connes' noncommutative
geometry \cite{book} induces  some very special
Yang-Mills and Higgs forces from the gravitational
force via particular generalized
coordinate transformations
\cite{tresch}\cite{grav}\cite{cc}. At the same time,
when acting on both the Dirac and Yang-Mills actions,
these coordinate transformations generate the
Yukawa terms, the Klein-Gordon action and the Higgs
potential with its spontaneous symmetry breaking.

The standard model
is in this very special class of Yang-Mills-Higgs
models as long as it fulfils the following requirements:
\begin{itemize}\item
Only weak isospin doublets and singlets, only colour
triplets and singlets occur in the fermion
representations.
\item
At least one neutrino is massless.
\item
The gauge symmetry of the gauge bosons that violate
parity is spontaneously broken by one complex Higgs
doublet. The corresponding weak gauge bosons,
$W^\pm,\ Z$ are massive and
$\rho:=m_W^2/(\cos^{2}\theta_w\,m_Z^{2})=1.$
\item
There is a Yang-Mills force, whose gauge group
commutes with weak isospin, hypercharge and with
the fermionic mass matrix, whose gauge group is
unbroken and whose couplings to fermions are
vectorial. Its gauge bosons, the gluons, are massless
\cite{reb}. 
\item
The hypercharges are such that one of
O'Raifeartaigh's reduction \cite{or}, namely to
$SU(2)\times U(1)\times SU(3)/\zz_3$ is possible. This
requirement permits the introduction of the
unimodularity condition \cite{ls}, the only ad hoc
feature in Connes' formulation.
\item
The coupling constants are constrained by
$g_2^2=g_3^2={\textstyle\frac{5}{3}}g_1^2
=3 \lambda. $
\end{itemize}
All these properties of the standard model are vital for
its geometrical interpretation: any version of the
standard model not satisfying one of these
requirements cannot be derived from gravity \`a la
Connes.

 We interpret the relations among the
coupling constants to hold at some energy scale
$\Lambda$. At this energy scale, the
noncommutativity of spacetime starts to be felt. Below
this scale, spacetime can be treated as a manifold.
As in grand unification we assume the big desert and
evolve the relations down to
$m_Z$ using the standard renormalisation flow. Under
this evolution $\lambda$ remains perturbative and
positive and yields a Higgs mass \cite{cc}\cite{bridge}
of
\bb m_H=182\pm 10\pm 7\ {\rm GeV}.\ee
The first error comes from the uncertainty in
$\Lambda=10^{13}-10^{17}$ GeV. The second is from
the present experimental uncertainty in the top mass,
$m_t=175\pm 6$ GeV.

Today a good number of reviews and books
\cite{rev} is available on the applications of Connes'
noncommutative geometry to gauge theories. I
recommend particularly the book by  J. M.
Gracia-Bond\'\i a, J. C. V\'arilly \& H. Figueroa
\cite{fgv} which is to appear soon.

This paper discusses the generalization of the
spin group to noncommutative geometry. It will give
us new conceptual insights into Connes' geometrical
description of all forces. For gravity, the commutative
case, this generalization reconciles Einstein's and
Cartan's view points. A natural, still commutative
extension unifies gravity and electromagnetism.
Finally the standard model will appear as the almost
simplest noncommutative extension. At the same time
we will discover a further subtle property of the
standard model, that is necessary in Connes'
description: O'Raifeartaigh's second reduction, by the
$\zz_2$. We will also get a partial answer to the old 
problem, what
symmetries of the fermion action are to be gauged.

\section{Lifting automorphisms to the Hilbert space}

Following Connes \cite{grav} consider a real, even
dimensional spectral triple given by:
\begin{itemize}\item
$\aa$, a real, associative algebra with unit 1 and
involution $\cdot^*$, $\aa$ is not necessarily
commutative,
\item
$\rho : \aa \longrightarrow \bbb(\hh)$, a faithful
representation of $\aa$ in terms of bounded operators
on a complex Hilbert space $\hh$,
\item
$\dd$, a self-adjoint, unbounded operator on $\hh$,
(`the Dirac operator'),
\item
$J$, an anti-unitary operator on $\hh$, (`the real
structure' or `charge conjugation'),
\item
$\chi $, a unitary operator on $\hh$, (`the chirality').
\end{itemize} 
The calibrating example is the commutative spectral
triple of a real, even dimensional Riemannian
spin-manifold discussed below. A certain number of
properties of this example are promoted to the
axioms of the general spectral triple:
$ J^2 =-1$ in 4 dimensions while $J^2=1$ in 0
dimensions, $\chi^2=1$, $J\chi=\chi J$, $\dd J=J\dd$, 
$\dd\chi=-\chi\dd$, $[\rho(a),J\rho(\tilde a)J^{-1}]=0$
for all $a,\tilde a\in\aa$, $[\dd,\rho(a)] $ is bounded
for all $a\in\aa$, $[[\dd,\rho(a)],J\rho(\tilde
a)J^{-1}]=0$ for all $a,\tilde a\in\aa$. This axiom is
called first order condition because in the calibrating
example it states that the genuine Dirac operator is a
first order differential operator. There are three more
axioms, that we do not spell out, orientability,
Poincar\'e duality and regularity.

 Let Aut($\aa$) denote the group of
automorphisms of $\aa$ and define its lift to the
Hilbert space $\hh$ to be the group ${\rm Aut}_\hh
(\aa)$.\hfil\break \\ \noindent
{\rm\bf Definition:} $
{\rm Aut}_\hh(\aa):=\{U\in {\rm End}(\hh),\ 
UU^* =U^*U=1,\ UJ=J U,\ U\chi =\chi U,
\hfil\break {} \hfil
 U \rho (a) U^{-1}\in
\rho (\aa)\  \forall a\in \aa\}.$\break \\ \noindent
 The first three properties say
that a lifted automorphism $U$ preserves probability,
charge conjugation and chirality. The fourth, called
{\it covariance property}, allows to define the
projection
$p:\ {\rm Aut}_\hh(\aa)\longrightarrow {\rm
Aut}(\aa)$ by
\bb (p(U))(a)=\rho ^{-1}(U\rho (a)U^{-1}).\ee
We will see that the covariance property is related to
the locality requirement of field theories.

 To
justify this definition we spell out the calibrating
example of a commutative geometry. For
concreteness let
$M$, (`spacetime') be a compact, 4-dimensional,
Riemannian spin-manifold. Its spectral triple is given
by:
\begin{itemize}
\item 
$\aa=\ccc^\infty(M)$, the {\it commutative} algebra of
complex valued functions on $M$ with complex
conjugation as involution,
\item
$\hh=\lll^2(\sss)$ is the Hilbert space of complex,
square integrable spinors $\psi $ on
$M$. In four dimensions spinors have four
components, $\psi (x)\in\cc^4,\ x\in M$ and we write
$\psi $ as column vector. The scalar product of two
spinors is defined by
\bb (\psi ,\psi ')=\int_M \psi ^*(x)\psi '(x)[\det g_{\mu
\nu }]^{1/2}\de ^4x,\ee where 
\bb g_{\mu \nu }=g\left( \,\frac{\pa}{\pa x^\mu }\,,
\,\frac{\pa}{\pa x^\nu }\,\right)
\ee  is the matrix of the Riemannian metric $g$ with
respect to the coordinates $x^\mu $,
$\mu =0,1,2,3.$ The representation is defined by
pointwise multiplication,
$(\rho(a)\,\psi )(x)=a(x)\psi (x),$ $x\in M,\ \psi
\in\lll^2(\sss)$.
\item
$\dd=\ddd$ is the genuine Dirac operator which we
write with respect to the chiral
$\gamma$ matrices,
\bb &\gamma ^0=\pp{0&-1_2\cr -1_2&0},\qq \gamma
^j={\textstyle\frac{1}{i}} 
\pp{0&\sigma _j\cr -\sigma _j&0},&\cr \cr  &\sigma
_1=\pp{0&1\cr 1&0},\qq\sigma _2=\pp{0&-i\cr
i&0},\qq\sigma _3=\pp{1&0\cr 0&-1},&\eee
\item
$J=C:=\gamma ^0\gamma ^2\circ\,{\rm complex\
conjugation}=\pp{0&-1&0&0\cr  1&0&0&0\cr
0&0&0&1\cr 0&0&-1&0}\,\circ\,{\rm c\ c}$,
\item
$\chi =\gamma _5=\gamma ^0\gamma ^1\gamma
^2\gamma ^3=
\pp{-1_2&0\cr 0&1_2}.$
\end{itemize} We have the following relations:
\bb \gamma ^a\gamma ^b+\gamma ^b\gamma
^a={\delta ^a}_b\,1_4,&a,b=0,1,2,3,&
\gamma ^{a*}=\gamma ^a,\\
 C^2=-1_4,& \gamma _5^2=1_4,& C\gamma _5=\gamma
_5C,\\ C\gamma ^a=-\gamma ^aC,&\gamma _5\gamma
^a=\gamma ^a\gamma _5&\\
\gamma ^{ab}:={\textstyle\frac{1}{2}} [\gamma
^a,\gamma ^b],&
\gamma ^{0j}=i\pp{-\sigma _j&0\cr 0&\sigma _j},&
\gamma ^{jk}=i\epsilon ^{jk\ell}
\pp{\sigma _\ell&0\cr 0&\sigma _\ell},\qq j,k=1,2,3.\ee
We recall that in the commutative case the spacetime
and its metric can be reconstructed from the
algebraic data of the spectral triple with its axioms.
The adjective `spectral' is motivated from Weyl's
theorem stating that the dimension of spacetime can
be recovered from the asymptotic behavior of the
ordered eigenvalues of the Dirac operator,
$...\lambda_n\le\lambda_{n+1}\le ...$, as they grow like
$n^{1/{\rm dim}\,M}$. The metric is
retrieved using Connes' formula for the geodesic
distance between two points,
$x,y\in M$,
\bb d(x,y)=\sup\{|a(x)-a(y)|;\ a\in\ccc^\infty(M),\ 
||[\ddd,\rho(a)]||\le 1\}.\label{dist}\ee
This equation suggests to take the Dirac operator as a
description of the metric even in a
noncommutative spectral triple.

In the commutative case, an algebra automorphism is
simply a diffeomorphism, Aut($\ccc^\infty(M)$) =
Diff($M$). We consider only diffeomorphisms
$\varphi$ close to the identity and we interpret
$\varphi $ as coordinate transformation, all our
calculations will be local, $M$ standing for one chart
on which the coordinate systems $x^\mu $ and $\tilde
x^{\tilde\mu }=(\varphi (x))^{\tilde\mu }$ are
defined. We will work out the local expression of a lift
of $\varphi $ to the Hilbert space of spinors. This lift
$U=L(\varphi )$ will depend on the metric and on the
initial coordinate system $x^\mu
$. 

In a first step we construct a group homomorphism
$\Lambda:{\rm Diff}(M)\rightarrow\,^MSO(4)$ onto
the group of local `Lorentz' transformations or
`Lorentz' gauge transformations, i.e. the group of
differentiable functions from spacetime into
$SO(4)$ with pointwise multiplication. Let
${(e^{-1}(x))^\mu }_a={(g^{-1/2}(x))^\mu }_a$ be the
inverse of the square root of the positive matrix of the
metric $g$ with respect to the initial coordinate system
$x^\mu $. Then the four vector fields $e_a$, $a=0,1,2,3$,
defined by
\bb e_a:= {(e^{-1})^\mu }_a\,\frac{\pa}{\pa x^\mu }\,
\ee 
give
 an orthonormal frame of the tangent bundle.
This frame defines a complete gauge fixing of the
Lorentz gauge group $^MSO(4)$
because it is the only orthonormal frame to have
symmetric coefficients ${(e^{-1})^\mu }_a$ with
respect to the coordinate system $x^\mu $. We call this
gauge the symmetric gauge for the coordinates
$x^\mu .$ Now let us perform a local change of
coordinates,
$\tilde x =\varphi (x)$. The holonomic frame with
respect to the new coordinates is related to the former
holonomic one by the inverse Jacobian matrix of
$\varphi $
\bb \,\frac{\pa}{\pa\tilde x^{\tilde \mu }}\,
=\,\frac{\pa x^\mu }{\pa\tilde x^{\tilde \mu }}\,
\,\frac{\pa}{\pa x^{ \mu }}\,={\left( \jj^{-1}\right)
^\mu }_{\tilde\mu }\,\frac{\pa}{\pa x^{\mu }},&&
{\left(
\jj^{-1}(\tilde x)\right) ^\mu }_{\tilde\mu
}=\,\frac{\pa x^\mu }{\pa\tilde x^{\tilde
\mu }}\,.\ee The matrix of the metric $g$ with respect
to the new coordinates reads,
\bb \tilde g_{\tilde\mu \tilde\nu }(\tilde x):=
g\!\!\left.\left(\,\frac{\pa}{\pa\tilde x^{\tilde \mu
}}\,,\,\frac{\pa}{\pa\tilde x^{\tilde \nu }}\,
\right)\right|_{\tilde x} =\left( \jj^{-1T}(\tilde x)
g(\varphi ^{-1}(\tilde x))\jj^{-1}(\tilde x)
\right) _{\tilde\mu \tilde\nu },\ee and the symmetric
gauge for the new coordinates $\tilde x$ is the new
orthonormal frame
\bb \tilde e_b=\hbox{$\tilde e^{-1\tilde\mu
}$}_b\,\frac{\pa  }{\pa\tilde x^{\tilde \mu
}}\,=\hbox{$\tilde g^{-1/2\,\tilde\mu
}$}_b{\jj^{-1\,\mu }}_{\tilde \mu
}\,\frac{\pa}{\pa x^{\mu }}\, ={\left(
\jj^{-1}\sqrt{\jj g^{-1}\jj^T}\right) ^\mu
}_b\,\frac{\pa}{\pa x^{
\mu}}\,.\ee New and old orthonormal frames are
related by a Lorentz transformation $\Lambda $,
$\tilde e_b={\Lambda ^{-1\,a}}_be_a$, with
\bb\left.\Lambda(\varphi)\right|_x
=\left.\sqrt{\jj^{-1T}g\jj^{-1}}\right|_{\varphi
(x)}\left.\jj
\right|_x\left.\sqrt{g^{-1}}\right|_x=\sqrt{\tilde
g}\jj\sqrt{g^{-1}}.\ee The dependence of $\Lambda
(\varphi )$ on the initial coordinates $x$ is natural,
\bb \left.\Lambda(\varphi)\right|_{\alpha (x)}=
\left.\Lambda(\alpha \varphi\alpha
^{-1})\right|_x.\label{alpha}\ee 
 This natural transformation under
a local diffeomorphism $\alpha $ allows to patch
together local expressions of the Lorentz
transformation in different overlapping charts.

If $M$ is flat and $x^\mu $ are `inertial' coordinates,
i.e. $g_{\mu \nu }={\delta ^\mu}_\nu $, and $\varphi $
is a local isometry then
$\jj(x)\in SO(4)$ for all
$x$ and
$\Lambda(\varphi ) =\jj$. In special relativity
therefore the symmetric gauge ties together Lorentz
transformations in spacetime with Lorentz
transformations in the tangent spaces.  

In general, if the coordinate transformation $\varphi
$ is close to the identity so is its Lorentz transformation
$\Lambda(\varphi )$ and it can be lifted to the spin
group,
\bb S: SO(4)&\longrightarrow& Spin(4)\cr 
\Lambda =\exp\omega &\longmapsto&\exp \left[
{\textstyle\frac{1}{4}} \omega _{ab}
\gamma ^{ab}\right] \label{spin}\ee with $\omega
=-\omega ^T\,\in so(4)$ and we can write the local
expression of the lift
 $L:{\rm Diff}(M)\rightarrow \,^MSpin(4)$,
\bb \left( L(\varphi )\psi \right) (\tilde
x)=\left.S\left(\Lambda (\varphi
)\right)\right|_{\varphi ^{-1}(\tilde x)}\psi
({\varphi^{-1}(\tilde x)}).\ee
$L$ is a locally bijective group homomorphism. For
any $\varphi $ close to the identity,
$L(\varphi )$ is unitary, commutes with charge
conjugation and chirality, satisfies the covariance
property, and $p(L(\varphi ))=\varphi $. Therefore
we have locally
\bb L({\rm Diff}(M))=\,^MSpin(4)\ \subset\  {\rm
Aut}_{\lll^2(\sss)}(\ccc^\infty (M)).\ee On the other
hand a local calculation shows that
\bb{\rm Aut}_{\lll^2(\sss)}(\ccc^\infty (M))={\rm
Diff}(M)\ltimes \,^MSpin(4).\ee 
The symmetric
gauge is a complete gauge fixing and is thereby
responsible for above reduction, the missing piece,
Diff($M$) being the set of $\alpha$'s in equation
(\ref{alpha}): 
\bb {\rm Diff}(M)\ltimes \,^MSO(4) \matrix{
{\rm sym.\ gauge\ fix.}\cr \longrightarrow\cr 
{}}{\rm Diff}(M)\ \matrix{
{L}\cr \longrightarrow\cr 
{}}\ \,^MSpin(4)\eee
This reduction follows Einstein's spirit in
the sense that the only arbitrary choice is the one of
the initial coordinate system
$x^\mu $ as will be illustrated in the next section.

Our computations are deliberately local. The global
picture is presented by Bourguignon
\& Gauduchon in reference \cite{bourg} of which this
section is a partial, pedestrian account.

\section{Einstein's dreisatz or let the flat metric
fluctuate}

The aim of this section is to reformulate Einstein's
derivation of general relativity,
\begin{itemize}\item Newton's law + Riemannian
geometry = Einstein's equations,
\end{itemize}
 in Connes' language of spectral triples.
As a by-product our lift $L$ will yield a self
contained introduction to Dirac's equation in a
gravitational field accessible to particle physicists.  

Einstein's starting point is the trajectory $x^\lambda 
(p)$ of a free particle in the flat spacetime of special
relativity. In inertial coordinates the dynamics is
given by 
\bb {\frac{\de^2x^\lambda}{\de p^2}}\,  =0.\ee 
Then Einstein goes to a uniformly accelerated system
$\tilde x^{\tilde \mu }$:
\bb {\,\frac{\de^2\tilde x^{\tilde
\lambda }}{\de p^2}\,}+ \hbox{$\tilde \Gamma
^{\tilde
\lambda }$}_ {\tilde
\mu \tilde \nu }(\tilde g) {\,\frac{\de\tilde x^{\tilde
\mu  }}{\de p}\,} {\,\frac{\de\tilde x^{\tilde \nu 
}}{\de p}\,}
 =0,\label{1}\ee 
 where a pseudo force appears. It is coded in the
Levi-Civita connection 
\bb\hbox{$\tilde \Gamma
^{\tilde
\lambda }$}_ {\tilde
\mu \tilde \nu }(\tilde g)={\textstyle\frac{1}{2}}
\tilde g^{\tilde \lambda\tilde \kappa}\left[
\,\frac{\pa}{\pa\tilde x^{\tilde \mu}}\,
\tilde g_{\tilde \kappa\tilde
\nu}+\,\frac{\pa}{\pa\tilde x^{\tilde \nu}}\,
\tilde g_{\tilde \kappa\tilde \mu}
-\,\frac{\pa}{\pa\tilde x^{\tilde \kappa}}\,
\tilde g_{\tilde \mu\tilde \nu}\right]\ee   
which
depends on the first partial derivatives of the matrix
$\tilde g_{\tilde \mu\tilde \nu}$ of the flat metric in
the new coordinates. The flat metric is hidden
in the initial, inertial coordinates.  Of course this
connection has vanishing curvature meaning that
this connection only describes pseudo forces. In a
first stroke Einstein relaxes this constraint and
declares the metric to be a dynamical variable. In a
second stroke Einstein looks for a suitable dynamics of
the metric which he finds completely determined by
the requirement that it be covariant under general
coordinate transformations, that it reproduces
Newton's law with the $1/r^2$ variation in the
non-relativistic limit (this is equivalent to looking
for second order differential equations) and that its
flat space limit be compatible with energy momentum
conservation. This dynamics is given by the Einstein
equation.

Connes' starting point is a free Dirac particle $\psi (x)$
in the flat spacetime of special relativity. In inertial
coordinates $x^\mu $ its dynamics is given by the
Dirac equation,
\bb \ddd\psi =i{\delta ^\mu }_a\gamma
^a{\,\frac{\pa}{\pa x^\mu }}\, \psi =0.\ee
We have written ${\delta ^\mu }_a\gamma
^a$ instead of $\gamma^\mu$ to stress that the
$\gamma$ matrices are $x$-independent. This Dirac
equation is covariant under Lorentz transformations.
Indeed if
$\varphi $ is a local isometry then 
\bb L(\varphi )\ddd L(\varphi )^{-1}=\tilde\ddd=
i{\delta ^{\tilde\mu} }_a\gamma
^a{\frac{\pa}{\pa \tilde x^{\tilde\mu} }}.\ee
To prove this special relativistic covariance one needs
the identity
$S(\Lambda )\gamma ^a S(\Lambda )^{-1}={\Lambda
^{-1\,a}}_b\gamma ^b$ for Lorentz transformations
$\Lambda \in SO(4)$ close to the identity. Now take a
general coordinate transformation
$\varphi $ close to the identity. A straight-forward
calculation \cite{sam} gives:
\bb L(\varphi )\ddd L(\varphi )^{-1}=\tilde\ddd=
i\hbox{$\tilde e^{-1\,\tilde\mu}$}_a\gamma^a\left[
\,\frac{\pa}{\pa\tilde x^{\tilde \mu}}\,+s(\tilde\omega
_{\tilde\mu})\right],\label{2}\ee
where
$\tilde e^{-1}=\sqrt{\jj \jj^T}$ is a symmetric matrix,
\bb s: so(4)&\longrightarrow& spin(4)\cr 
\omega &\longmapsto& 
{\textstyle\frac{1}{4}} \omega _{ab}
\gamma ^{ab} \ee
is the Lie algebra isomorphism corresponding to the
lift (\ref{spin}) and
\bb \tilde
\omega_{\tilde\mu}(\tilde
x)=\left.\Lambda\right|_{\varphi^{-1}(\tilde x)}
\pa_{\tilde\mu}
\left.\Lambda^{-1}\right|_{\tilde x},&&
\tilde
\pa_{\tilde\mu}:=
\,\frac{\pa}{\pa\tilde x^{\tilde \mu}}\,.\ee
The `spin connection' $\tilde \omega$ is identical to
the Levi-Civita connection $\tilde \Gamma$, the only
difference being that the latter is expressed with
respect to the holonomic frame $\pa_{\tilde\mu}$,
while the former is written with respect to the
orthonormal frame $\tilde e_a=\hbox{$\tilde
e^{-1\,\tilde\mu}$}_a\pa_{\tilde\mu}$.
We recover
 the well known explicit expression
\bb \tilde {\omega^a}_{b\tilde \mu}(\tilde e)=
{\textstyle\frac{1}{2}}\left[(\tilde \pa_{\tilde
\beta}\hbox{$\tilde e^a$}_{\tilde \mu})-
(\tilde \pa_{\tilde
\mu}\hbox{$\tilde e^a$}_{\tilde \beta})+
\hbox{$\tilde e^m$}_{\tilde \mu}(\tilde
\pa_{\tilde\beta  }
\hbox{$\tilde e^m$}_{\tilde
\alpha})\hbox{$\tilde e^{-1\,\tilde\alpha}$}_a
\right]\hbox{$\tilde e^{-1\,\tilde
\beta}$}_b\ -\ [a \leftrightarrow b]\ee
 of the spin
connection in terms of the first derivatives of 
$\hbox{$\tilde e^a$}_{\tilde \mu}= {\sqrt{\tilde
g}^a}_{\tilde
\mu}.$ Again the spin connection has zero curvature
and the first stroke relaxes this
constraint. But now equation (\ref{2}) has an
advantage over its analogue (\ref{1}). Thanks to
Connes' distance formula (\ref{dist}), the metric can
be read explicitly in (\ref{2}) from the matrix of
functions
$\hbox{$\tilde e^{-1\,\tilde\mu}$}_a$ while in
(\ref{1}) only the first derivatives of the metric are
present. We are used to this nuance from
electromagnetism where the classical particle feels
the force while the quantum particle feels the
potential. In Einstein's approach the zero connection
fluctuates, in Connes' approach the flat metric
fluctuates. This means that the constraint $\tilde
e^{-1}=\sqrt{\jj
\jj^T}$ is relaxed and $\tilde e^{-1} $ now is an
arbitrary symmetric matrix depending smoothly
on $\tilde x$.

 The second stroke, the covariant dynamics
for the new class of Dirac operators $\tilde \ddd$, is
due to Chamseddine \& Connes \cite{cc}. This is the
celebrated spectral action.
 The beauty of their
approach to general relativity is that it works
precisely because the Dirac operator $\tilde \ddd$
plays two roles simultaneously, it defines the
dynamics of matter and it parameterizes the set of all
Riemannian metrics. For a discussion of the
transformation passing from the metric to the Dirac
operator I recommend the article \cite{lr} by Landi
\& Rovelli. 

The starting point of Chamseddine \& Connes is the
simple remark that the spectrum of the Dirac operator
is invariant under diffeomorphisms interpreted as
general coordinate transformations. From
$\tilde \ddd\chi=-\chi\tilde \ddd$ we know that the
spectrum of
$\tilde \ddd$ is even. We may therefore consider only
the spectrum of the positive operator $\tilde
\ddd^2/\Lambda^2$ where we have divided by a fixed
arbitrary energy scale to make the spectrum
dimensionless. If it was not divergent the trace $\t
\tilde \ddd^2/\Lambda^2$ would be a general
relativistic action functional. To make it convergent,
take a differentiable function
$f:\rr_+\rightarrow\rr_+$ of sufficiently fast
decrease such that the action
\bb S_{CC}:=\t f(\tilde \ddd^2/\Lambda^2)\ee
converges. It is still a diffeomorphism invariant
action. Using the heat kernel expansion it can be
computed asymptotically:
\bb S_{CC}=
\int_M
[\Lambda_c-{\textstyle\frac{m_P^2}{16\pi}}R
+a(5\,R^2-8\,{\rm Ricci}^2-7\,{\rm
Riemann}^2)]\,\sqrt{\det g_{\mu\nu}}\de^4x \,+\,
O(\Lambda^{-2}),\ee where the cosmological constant
is $\Lambda_c=
{\textstyle\frac{f_0}{4\pi^2}}\Lambda^4$, the Planck
mass is
$m_P^2={\textstyle\frac{f_2}{3\pi}}\Lambda^2$ and 
$a={\textstyle\frac{f_4}{5760\pi^2}}$. The
Chamseddine-Connes action is universal in the sense
that the `cut off' function $f$ only enters through its
first three `moments', $f_0:=\int_0^\infty uf(u)\de u$, 
$f_2:=\int_0^\infty f(u)\de u$ and $f_4=f(0)$. Thanks
to the curvature  square terms the
Chamseddine-Connes action is positive and has
minima. For instance the 4-sphere with a radius of
$(11f_4)^{1/2}(90\pi
(1-(1-11/15\,f_0f_4f_2^{-2})^{1/2}))^{-1/2}$
 times the Planck length is a ground state. This
minimum breaks the diffeomorphism group
spontaneously down to the isometry group $SO(5)$.
The little group consists of those lifted
automorphisms that commute with the Dirac operator
$\tilde \ddd$. Let us anticipate that the spontaneous
symmetry breaking of the Higgs mechanism will be a
mirage of this gravitational break down. I must admit
that it took me four years to understand what Connes
meant by this gravitational symmetry breaking.
Physically it seems to regularize the initial
cosmological singularity. 

We close this section with a side remark. We noticed
that the matrix $\hbox{$\tilde
e^{-1\,\tilde\mu}$}_a$ in equation (\ref{2}) is
symmetric. A general, not necessarily symmetric
matrix
$\hbox{$\hat e^{-1\,\tilde\mu}$}_a$ can be obtained
from a general Lorentz transformation
$\Lambda\in\,^MSO(4)$:
\bb \hbox{$\tilde e^{-1\,\tilde\mu}$}_a {\Lambda^a}_b
=\hbox{$\hat e^{-1\,\tilde\mu}$}_b,\ee
which is nothing but the polar decomposition of the
matrix $\hat e^{-1}$.

\section{The gauge dreisatz or let the metric fluctuate
again}

At this point we are reminded of a dreisatz very similar
to the above one in Connes' formulation:
\begin{itemize}\item free Schr\"odinger equation +
gauge invariance = Maxwell's equations.
\end{itemize}
Indeed the free Schr\"odinger equation is covariant
under phase transformations of the wave function,
$\psi\mapsto \exp (i\theta_V)\psi$ for real constant
$\theta_V$. In the first stroke, we want to enlarge the
$U(1)$ group of phase transformations to the gauge
group $^MU(1)$. This is possible if we introduce the
real gauge connections $A_\mu$ and replace the 
partial derivatives
$\pa_\mu$  in the
free Schr\"odinger equation by the covariant 
derivatives
$\pa_\mu+(iq/\hbar) A_\mu$ where $q$ is the
electric charge of the Schr\"odinger particle that
 loses its freedom. From now on we put $\hbar=1$.
In the second stroke we want to promote the
gauge connection to a dynamical variable. If we want
the dynamics to be gauge covariant and to be given by
second order differential equations (because of the
$1/r^2$ variation in Coulomb's law) then the answer
is unique: the Maxwell equations. 

In Connes' formulation the group of $U(1)$ gauge
transformations appears naturally, it is the group of
unitaries,
\bb U(\aa):=\left\{ u\in \aa,\ uu^*=u^*u=1\right\}
,\ee
of the algebra $\aa=\ccc^\infty(M)$. It is tempting to
try and repeat the gauge dreisatz with the Dirac
equation. However the representation of a unitary $u$
on the Hilbert space of spinors $\hh=\lll^2(\sss)$ does
not commute with charge conjugation. The reason is
clear, the 4-component spinor $\psi $ contains
particles and antiparticles. If particles transform with
$u$ then antiparticles must transform with $u^*$
because they have opposite electric charge. To
disentangle particles and antiparticles, Connes doubles
the fermions, $\hh_t=
\hh\op\hh=\lll^2(\sss)\ot\cc^2$, and defines a new
spectral triple:
\bb \aa=\ccc^\infty(M)\owns a,&
\hh_t=\lll^2(\sss)\ot\cc^2\owns\psi _t=\pp{\psi \cr
\psi ^c},&
\rho _t(a)=\pp{a1_4&0\cr 0&\bar a1_4},\\
\dd_t=\pp{\ddd&0\cr 0&\ddd},&J_t=\pp{0&-1&0&0\cr 
1&0&0&0\cr 0&0&0&1\cr 0&0&-1&0}\ot\pp{0&1\cr
1&0}\circ\,{\rm c\ c},&\chi _t=\gamma
_5\ot\pp{1&0\cr 0&1}.\ee
We anticipate that $\psi ^c$ is not a new degree of
freedom but we will make $\psi ^c$ the antiparticle of
$\psi $ at the end of the day by imposing $J_t\psi
_t=\psi _t.$ This disentangling of particles and
antiparticles is close to Dirac's spirit who
reinterprets the antiparticles as holes. 

Now Connes defines a second lift into the group of
generalized automorphisms ${\rm Aut}_{\hh_t}
(\aa)$,
\bb \ell:U(\aa)&\longrightarrow&
{\rm Aut}_{\hh_t} (\aa)\cr 
u_V=\exp(i\theta_V (x))&\longmapsto&
\ell(u_V)=\rho_t(u_V)J_t\rho_t(u_V)J_t^{-1}=:U_V
.\ee
Note that $p(\ell(u_V))=1$ for
every unitary $u_V$.
Note also that without fermion doubling, $\ell$ alone 
would be already trivial, $\ell(u_V)=1$. Let us put both
lifts together,
\bb (L,\ell):{\rm Aut}(\aa)\ltimes U(\aa)
&\longrightarrow&\,^M\left(Spin(4)\times
U(1)_V\right)\ 
\subset\ {\rm Aut}_{\hh_t} (\aa)\cr 
(\varphi,u_V)&\longmapsto&
\left(L(\varphi),\ell(u_V)\right)\eee 
\bb\left(\left(L(\varphi),\ell(u_V)\right)\psi_t\right)
(\tilde x)=
\left.S\left(\Lambda (\varphi
)\right)\right|_{\varphi ^{-1}(\tilde x)}
\ot\pp{u_V^2({\varphi^{-1}(\tilde x)})&0\cr 
0&\bar u_V^2({\varphi^{-1}(\tilde x)})}
\psi_t({\varphi^{-1}(\tilde x)}).\ee
Note the exponents two coming from fermion
doubling.  What fluctuations do we get now if we start
again from the free Dirac operator $\dd_t$ with
$\ddd=i{\delta ^\mu }_a\gamma ^a{\,\frac{\pa}{\pa
x^\mu }}$ 
\bb\left(L(\varphi),\ell(u_V)\right)\dd_t
\left(L(\varphi),\ell(u_V)\right)^{-1}=\tilde\dd_t=
\pp{\tilde \ddd&0\cr 0&{C\tilde\ddd C^{-1}}}?\ee
As before, a straight-forward calculation yields the
covariant derivative:
\bb \tilde\ddd=
i\hbox{$\tilde e^{-1\,\tilde\mu}$}_a\gamma^a\left[
\tilde\pa_{\tilde \mu}+s(\tilde\omega
_{\tilde\mu})-2i\tilde A_{\tilde\mu}\right]. \ee
The Maxwell connection 
\bb\tilde A_{\tilde\mu}={\textstyle\frac{1}{i}} 
u_V\tilde\pa_{\tilde \mu }u_V^{-1}
=\,\frac{\pa\theta_V}{\pa\tilde
x^{\tilde \mu}}\ee
acts on particles $\psi $  as
\bb 2\gamma ^{\tilde \mu }\tilde A_{\tilde \mu }=(-i)
\rho (u_V)\left[ \ddd, \rho (u_V)^{-1}\right]
+C(-i)
\rho (u_V)\left[ \ddd, \rho (u_V)^{-1}\right]
C^{-1}
. \ee
Comparing with the gauge dreisatz before we see that
the electric charge is quantized, it admits only two
values, $q=-2$ in $\tilde\ddd$  and
$q=+2$ in $C\tilde\ddd C^{-1}$.
The Maxwell connection has zero field strength, of
course. The first stroke relaxes the constraints of
vanishing curvature and of vanishing field strength.
The second stroke is again the spectral action and it
unifies gravity and electrodynamics:
\bb S_{CC}&=&\t f(\tilde \ddd^2/\Lambda^2)\cr\cr  &=&
\int_M
[\Lambda_c-{\textstyle\frac{m_P^2}{16\pi}}R
+a(5\,R^2-8\,{\rm Ricci}^2-7\,{\rm
Riemann}^2)\cr
&&\qq\qq\qq\qq\qq\qq\qq\qq\qq\qq\qq
+{\textstyle\frac{1}{4g^2}}F_{\mu
\nu }^*
 F^{\mu \nu }
]\,\sqrt{\det g_{\mu\nu}}\de^4x \,+\,
O(\Lambda^{-2}),\ee 
where the electric coupling  constant
is $g^2={\textstyle\frac{6\pi ^2}{f_4}} $.

\section{A second fermion doubling, a third
fluctuation}

Consider the spin cover
$p:Spin(4)\rightarrow SO(4)$. Every element close to
the identity upstairs (in $Spin(4)$) can be obtained by
lifting an element from downstairs. This is also the
case for our initial spectral triple, 
$p: {\rm Aut}_{\hh}(\ccc^\infty (M))
={\rm Diff}(M)\ltimes\,^MSpin(4)
\rightarrow {\rm Aut}(\ccc^\infty (M))={\rm
Diff}(M)$. 
After the fermion doubling however, this is no longer
true. Indeed a local calculation gives
\bb{\rm Aut}_{\hh_t}(\ccc^\infty (M))&=&{\rm
Diff}(M)\ltimes \,^M\left(Spin(4)\times
U(1)_V\times U(1)_A\right),\ee
see figure.
 \begin{figure}[hbt]
\epsfxsize=7.1cm
\hspace{4cm}
\epsfbox{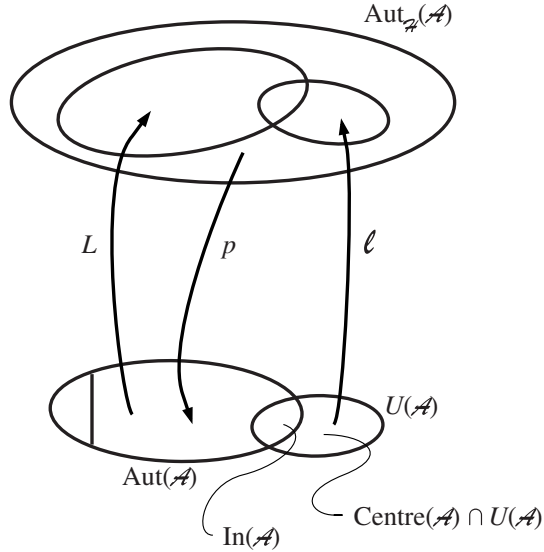}
\caption{Lifting automorphisms and unitaries}
\end{figure}

Let us denote the elements of $^M\left(U(1)_V\times
U(1)_A\right)$ by \hfil\break
 $(U_V,U_A)=(\exp(i\theta_V(x)),
\exp(i\theta_A(x)))$.
Their action on a spinor $\psi_t(x)$ is given by
\bb R(U_V,U_A)=\pp{\exp
\left\{2i\left(\theta_V1_4+\theta_A\gamma_5
\right)\right\}&0\cr 0&
\exp\left\{ -2
i\left(\theta_V1_4+\theta_A\gamma_5\right)\right\}
},\ee
\bb \exp
\left\{\pm 2i\left(\theta_V1_4+\theta_A\gamma_5
\right)\right\}=
\exp
\left\{\pm 2i\left(\theta_V-\theta_A
\right)\right\}\,\frac{1-\gamma_5}{2}\,+\,
\exp
\left\{\pm 2i\left(\theta_V+\theta_A
\right)\right\}\,\frac{1+\gamma_5}{2}
\,.\ee
While the Maxwell gauge transformation $U_V$
($\cdot_V$ for vectorial) comes from a unitary,
$R(U_V,1)=\ell(u_V)$, the chiral transformation
$U_A$ ($\cdot_A$ for axial) is an uninvited guest.
Connes writes him a letter of invitation by doubling
fermions once again. He defines a new
spectral triple:
\bb&&
\aa_t=\ccc^\infty(M)\ot\left(\cc_L\op\cc_R\right)
\owns (a_L,a_R),\\ \nonumber\\[2mm] &&
\hh_t=\lll^2(\sss)\ot\cc^4\ \owns\psi _t=\pp{\psi_L
\cr \psi_R \cr\psi^c_L\cr 
\psi _R^c},\qq
\rho _t(a_L,a_R)=\pp{a_L1_4&0&0&0
\cr 0&a_R1_4&0&0\cr 
0&0&\bar a_L1_4&0\cr 
0&0&0&\bar a_R1_4},\\ \nonumber\\[2mm] &&
\dd_t=\pp{\ddd&0&0&0\cr
0&\ddd&0&0\cr 
0&0&\ddd&0\cr 
0&0&0&\ddd},\\ \nonumber\\[2mm] &&
J_t=\pp{0&-1&0&0\cr  1&0&0&0\cr
0&0&0&1\cr 0&0&-1&0}\ot
\pp{0&0&1&0\cr
0&0&0&1\cr 
1&0&0&0\cr 
0&1&0&0}\circ\,{\rm
c\ c},\\ \nonumber\\[2mm] &&
\chi _t=\gamma _5\ot\pp{-1&0&0&0\cr
 0&1&0&0\cr 
0&0&-1&0\cr 
0&0&0&1}.\ee
This second doubling is to disentangle
left- and right-handed fermions and it is an old
friend from Euclidean Lagrangian field theory with
chiral fermions. Like the first doubling is does not
add new degrees of freedom: at the end of the day {\it
and} after passage to the Minkowskian, half of the
fermions are projected out by imposing $\chi _t\psi
_t=\psi _t.$

Real, even spectral triples are natural in the sense that
the tensor product of two triples $(\aa_i,\hh_i,\dd_i,
J_i,\chi _i)$, $i=1,2$ 
of even dimensions
$d_1$ and
$d_2$ is a triple $(\aa_t,\hh_t,\dd_t,J_t,\chi _t)$ of
dimension $d_1+d_2$. This tensor product is defined by
\bb &&\aa_t=\aa_1\ot\aa_2,\qq
\hh_t=\hh_1\ot\hh_2,\cr 
&&\dd_t=\dd_1\ot 1_2\ +\ \chi _1\ot\dd_2,\cr 
&&J_t=J_1\ot J_2,\qq\chi _t=\chi _1\ot\chi _2.\eee
The second obvious choice for the Dirac operator,
$\dd_1\ot\chi _2\ +\ 1_1\ot\dd_2$, is unitarily
equivalent to the first one. After this second doubling,
apparently we are in presence of such a tensor
product: the first triple describes 4-dimensional
spacetime,
\bb\left(
\ccc^\infty(M),\lll^2(\sss),\ddd,C,\gamma_5\right),
\ee
 the second describes the 0-dimensional
two-point space, 
\bb\aa_f= \cc_L\op\cc_R\owns
(a_L,a_R),\qq
\hh_f=\cc^4,\qq
\rho_f(a_L,a_R)=\pp{a_L&0&0&0\cr 
0&a_R&0&0\cr 
0&0&\bar a_L&0\cr 
0&0&0&\bar a_R},\ee
\bb\dd_f=0,\qq
J_f=\pp{0&0&1&0\cr 
0&0&0&1\cr 
1&0&0&0\cr 
0&1&0&0}\circ {\rm c\ c},\qq
\chi _f=\pp{-1&0&0&0\cr 0&1&0&0\cr 0&0&-1&0\cr
0&0&0&1}.\ee
 We have indicated the second triple by
the subscript $\cdot_f$ (for finite) rather than by
$\cdot_2$. Since the Dirac operator $\dd_f$ vanishes
the two points are separated by an infinite distance
according to Connes' distance formula (\ref{dist}). We
want to make this distance finite. The most general
finite Dirac operator commuting with $J_f$ and
anticommuting with $\chi_f$ is:
\bb \dd_f =\pp{0&m&0&0\cr \bar
m&0&0&0\cr 
0&0&0&\bar m\cr0&0&m&0 },&&m\in\cc.\ee 
Now the distance
between the points is $1/|m|$. On the other hand 
$\dd_t=\ddd\ot1_4\ +\
\gamma_5\ot\dd_f$ is precisely the free massive
Euclidean Dirac operator. 

The tensor product of the above triples describes the
two-sheeted universe, with two flat sheets at
constant distance. We are eager to see this free
Dirac operator $\dd_t$ fluctuate. 
\bb\left(L(\varphi),\ell(u_L,u_R)\right)\dd_t
\left(L(\varphi),\ell(u_L,u_R)\right)^{-1}=
\pp{\tilde \ddd_L&\Phi\gamma_5 &0&0\cr 
\bar \Phi\gamma_5 &\tilde \ddd_R&0&0\cr 
0&0&C\tilde \ddd_LC^{-1}&\bar \Phi\gamma_5 \cr 
0&0&\Phi\gamma_5 &C\tilde
\ddd_RC^{-1}},\label{dt}\ee with two gauge bosons
\bb
\tilde A_{L\tilde\mu}={\textstyle\frac{1}{i}} 
u_L\tilde\pa_{\tilde \mu }u_L^{-1}
=\,\frac{\pa\theta_L}{\pa\tilde
x^{\tilde \mu}},\qq
\tilde A_{R\tilde\mu}={\textstyle\frac{1}{i}} 
u_R\tilde\pa_{\tilde \mu }u_R^{-1}
=\,\frac{\pa\theta_R}{\pa\tilde
x^{\tilde \mu}},\ee
their corresponding covariant derivatives,
\bb 
 \tilde\ddd_L=
i\hbox{$\tilde e^{-1\,\tilde\mu}$}_a\gamma^a\left[
\tilde\pa_{\tilde \mu}+s(\tilde\omega
_{\tilde\mu})-2i\tilde A_{L\tilde\mu}\right],\qq
 \tilde\ddd_R=
i\hbox{$\tilde e^{-1\,\tilde\mu}$}_a\gamma^a\left[
\tilde\pa_{\tilde \mu}+s(\tilde\omega
_{\tilde\mu})-2i\tilde A_{R\tilde\mu}\right],\ee
and as star guest: the Higgs boson
\bb \Phi =m+u_L^2mu_R^{-2}.\ee
Connes has generalized the exterior derivative to
arbitrary spectral triples and in his sense $\Phi $ is a
connection 1-form describing parallel transport
between the two sheets. Its curvature vanishes. In the
first stroke the metric, the two gauge bosons
and the Higgs are promoted to dynamical variables
with arbitrary curvature. According to Connes'
distance formula (\ref{dist}), this new kinematics now
describes two sheets with arbitrary but identical
metric and with variable separation $1/|\Phi| $. In the
second stroke the spectral action produces (in addition
to the familiar dynamics of metric and gauge bosons)
the Klein-Gordon action for the Higgs, covariant with
respect to the gauge bosons, and the quartic Higgs
potential, that breaks
$U(1)_L\times U(1)_R$ spontaneously down to
$U(1)_A$. The Yukawa couplings, necessary to allow
us to view the fermion mass $m$ as generated by this
spontaneous symmetry breaking, stem from the
fluctuation total Dirac operator $\dd_t$, equation
(\ref{dt}), and they have a natural interpretation as
covariant derivative with respect to a transport
between the two sheets. The Higgs is celebrated as star
guest because he was not invited to this party, a party,
that rehabilitates the entire Higgs mechanism. 

Physically, the model
contains two gauge bosons, a massive one with axial
couplings and a massless one with vector couplings.
For this reason Connes \& Lott \cite{cl} called this
model `chiral electrodynamics'. Its initial setting is
that of a left-right symmetric model with left- and
right-handed gauge bosons. However the Higgs
sector, on which there is no handle in Connes' setting,
decides that the eigenstates of the mass matrix of the
gauge bosons have vector and axial couplings.
Consequently parity is not broken spontaneously. This
is a general feature in noncommutative geometry
\cite{leri}.

At this point one remark is in order. Our
initial motivation was a certain balance between
automorphisms and unitaries on the one side and
lifted automorphisms on the other side. Now we have a
new phenomenon, there are automorphisms close to
the identity that cannot be lifted, see figure. Indeed
locally, Aut($\aa_t$)\,=\,Diff($M)\times $Diff($M)\,
\owns\,(\varphi _L,\varphi _R).$ However only
automorphisms satisfying $\varphi _L=\varphi _R$
can be lifted to the Hilbert space. This phenomenon
guarantees that the massive Dirac action remains local
in the sense of field theory, i.e. the Lagrangian only
contains products of fields and of a finite number of
their derivatives at the same spacetime point.

The finite spectral triple of the two-point space still
has one short coming, it does not satisfy the first order
axiom. There
are two ways to fix this problem:
We may minimally modify the representation such
that it becomes vector-like in the antiparticle sector,
e.g.,
\bb \aa_f=\cc\op\cc\owns (a,b),\qq
\rho_f(a,b)=\pp{a&0&0&0\cr 
0&\bar b&0&0\cr 
0&0&b&0\cr 
0&0&0& b}.\ee
Then after the fluctuation of the metric, the charge
quantisation is less restrictive, $q=-2,0,+2$. This
possibility is realized in the standard model in the
lepton sector. The second possibility is to enlarge
the algebra by adding a third factor, for example
another $\cc$, and represent it vectorially,
\bb\aa_f= \cc\op\cc\op\cc\owns
(a,b,c),\qq
\rho_f(a,b,c)=\pp{a&0&0&0\cr 
0&b&0&0\cr 
0&0&\bar c&0\cr 
0&0&0&\bar c}.\ee
In the standard model, $c$ will be the colour. We have
noticed above that there may be automorphisms close
to the identity that cannot be lifted to the Hilbert
space. Now, this last spectral triple has
unitaries close to the identity that are lifted to the
identity. Indeed, $U(\aa_f)=U(1)^3$, but
$\ell(U(\aa_f))=U(1)^2$.  Also note that this last
spectral triple alone does not satisfy the Poincar\'e
duality, it must be accompanied by another triple, e.g.
the former one. 

\section{The standard model}

So far our spectral triples were commutative. Connes'
geometry never uses this property and develops its
full power in the noncommutative case. For instance,
close to the identity, Aut$(\hhh)=U(\hhh)=SU(2)$ for
the noncommutative algebra $\hhh$ of quaternions
and there is no need to introduce a central extension
$\ell$. From the physical point of view
noncommutative triples are welcome because they
offer us spontaneously broken non-Abelian
Yang-Mills theories. For these applications, it is
sufficient to consider only mildly noncommutative
triples: tensor products of the infinite dimensional
commutative triple describing 4-dimensional
spacetime with a finite dimensional noncommutative
triple, `the internal space'. We call such tensor
products almost commutative spaces. Madore \cite{kk}
uses the word Kaluza-Klein spaces because they have
the geometrical interpretation of a direct product of a 
4-dimensional manifold with a discrete point set
\cite{ikm} as the two sheeted universe. 

Only very few Yang-Mills-Higgs models can be
formulated as almost commutative geometries
\cite{versus} and can thereby be viewed as fluctuations
of general relativity. We cannot believe that it is pure
coincidence that the intricate standard model of
electro-weak and strong forces is among these very
few models. The weak force breaks parity, however
parity cannot be broken spontaneously in Connes'
approach \cite{leri} and must therefore be broken
explicitly in the finite dimensional triple. The
simplest way to do so is to choose
$a_L$ and $a_R$ in algebras of different dimensions,
say $\dim \aa_L>\dim \aa_R$. As immediate
consequence parity will then be maximally broken by
purely left-handed gauge bosons as in the standard
model. 

 Here is its internal space:
The algebra is chosen as to reproduce 
$SU(2)\times U(1)\times SU(3)$ as subgroup of
$U(\aa)$,
\bb \aa_f=\hhh\op\cc\op
M_3(\cc)\,\owns\,(a,b,c).\ee The internal Hilbert
space is copied from the Particle Physics Booklet
\cite{data} as given in equations (\ref{hl}),(\ref{hr}),
\bb \hh_L&=&
\left(\cc^2\ot\cc^N\ot\cc^3\right)\ \op\ 
\left(\cc^2\ot\cc^N\ot\cc\right), \\
\hh_R&=&\left(\cc^N\ot\cc^3\right)\ 
\op\ \left(\cc^N\ot\cc^3\right)\ 
\op\ \left(\cc\ot\cc^N\ot\cc\right).\ee
 In each summand, the first factor
denotes weak isospin doublets or singlets, the second
denotes
$N$ generations, $N=3$, and the third denotes colour
triplets or singlets.
Let us choose the following basis
of 
$\hh_f=\hh_L\op\hh_R\op\hh^c_L\op\hh^c_R
=\cc^{90}$: 
\bb
& \pp{u\cr d}_L,\ \pp{c\cr s}_L,\ \pp{t\cr b}_L,\ 
\pp{\nu_e\cr e}_L,\ \pp{\nu_\mu\cr\mu}_L,\ 
\pp{\nu_\tau\cr\tau}_L;&\cr \cr 
&\matrix{u_R,\cr d_R,}\qq \matrix{c_R,\cr s_R,}\qq
\matrix{t_R,\cr b_R,}\qq  e_R,\qq \mu_R,\qq 
\tau_R;&\cr  \cr 
& \pp{u\cr d}^c_L,\ \pp{c\cr s}_L^c,\ 
\pp{t\cr b}_L^c,\ 
\pp{\nu_e\cr e}_L^c,\ \pp{\nu_\mu\cr\mu}_L^c,\ 
\pp{\nu_\tau\cr\tau}_L^c;&\cr\cr  
&\matrix{u_R^c,\cr d_R^c,}\qq 
\matrix{c_R^c,\cr s_R^c,}\qq
\matrix{t_R^c,\cr b_R^c,}\qq  e_R^c,\qq \mu_R^c,\qq 
\tau_R^c.&\eee
It is the current eigenstate basis, the representation
$\rho_f$ acting on
$\hh_f$ by
\bb \rho_f(a,b,c):= 
\pp{\rho_{L}&0&0&0\cr 
0&\rho_{R}&0&0\cr 
0&0&{\bar\rho^c_{L}}&0\cr 
0&0&0&{\bar\rho^c_{R}}}\ee
with
\bb\rho_{L}(a):=\pp{
a\ot 1_N\ot 1_3&0\cr
0&a\ot 1_N&},\qq
\rho_{R}(b):= \pp{
b 1_N\ot 1_3&0&0\cr 0&\bar b 1_N\ot 1_3&0\cr 
0&0&\bar
b1_N},
\ee\bb 
  \rho^c_{L}(b,c):=\pp{
1_2\ot 1_N\ot c&0\cr
0&\bar b1_2\ot 1_N},\qq
\rho^c_{R}(b,c) := \pp{
1_N\ot c&0&0\cr 0&1_N\ot c&0\cr
0&0&\bar b1_N}.   
\ee
At this point we understand why only isospin
doublets and singlets and colour triplets and singlets
can be used in the fermionic representation:
all other irreducible group representations cannot be
extended to algebra representation. While the
tensor product of two group representations is again a
group representation, the tensor product of two
algebra representations is not an algebra
representation.
 The
apparent asymmetry between particles and
antiparticles -- the former are subject to weak, the
latter to strong interactions -- disappears after
application of the lift $\ell$ with
\bb J_f=\pp{0&1_{15N}\cr 1_{15N}&0}\circ 
\ {\rm complex\ conjugation}.\ee
 For the sake of
completeness, we record the chirality as matrix
\bb \chi_f=\pp{-1_{8N}&0&0&0\cr 0&1_{7N}&0&0\cr
 0&0&-1_{8N}&0\cr 0&0&0&1_{7N} }.\ee
The
internal Dirac operator
\bb \dd_f=\pp{0&\mm&0&0\cr 
\mm^*&0&0&0\cr 
0&0&0&\bar\mm\cr 
0&0&\bar\mm^*&0}\ee
contains the fermionic mass matrix of the standard
model,
\bb\mm=\pp{
\pp{1&0\cr 0&0}\ot M_u\ot 1_3\,+\,
\pp{0&0\cr 0&1}\ot M_d\ot 1_3
&0\cr
0&\pp{0\cr 1}\ot M_e},\ee
with
\bb M_u:=\pp{
m_u&0&0\cr
0&m_c&0\cr
0&0&m_t},&& M_d:= C_{KM}\pp{
m_d&0&0\cr
0&m_s&0\cr
0&0&m_b},\\[2mm] M_e:=\pp{
m_e&0&0\cr
0&m_\mu&0\cr
0&0&m_\tau}.&&\ee
From the booklet we know that all indicated fermion
masses are different from each other and that the
Cabibbo-Kobayashi-Maskawa matrix  $C_{KM}$ is
non-degenerate in the sense that  no quark is
simultaneously mass and weak current eigenstate.

We note that Majorana masses are forbidden because
of the axiom $\dd_f\chi_f=-\chi_f\dd_f.$ At least one
neutrino must be without a right-handed piece in
order to fulfil the Poincar\'e duality which for a
finite dimensional spectral triple states that the
intersection form
\bb \cap_{ij}:=\t
\left[\chi_f\,\rho_f(p_i)\,J_f\rho_f(p_j)J_f^{-1}
\right]\ee
 must be non-degenerate. The $p_j$ are a set of
minimal projectors of $\aa_f$. The standard model
has  three minimal
projectors, 
\bb p_1=(1_2,0,0),\qq p_2=(0,1,0), \qq p_3=
\left(0,0,\pp{1&0&0\cr 0&0&0\cr 0&0&0}\right)\ee
and the intersection form with three purely
left-handed neutrinos,
\bb \cap=-6\pp{0&1&1\cr 1&-1&-1\cr 1&-1&0},\ee 
is non-degenerate. However if we add three
right-handed neutrinos to the standard model,
massive or not, then the intersection form,
\bb \cap=-6\pp{0&1&1\cr 1&-2&-1\cr 1&-1&0},\ee 
is degenerate and Poincar\'e duality fails.

The first order axiom,
$[[\dd_f,\rho_f(a)],J_f\rho_f(\tilde a)J_f^{-1}]=0$ for
all $a,\tilde a\in\aa_f$ requires a
gauge group that commutes with the electro-weak
interactions and with the fermionic mass matrix and
whose fermion representation is vectorial \cite{reb}.

The fluctuation of the free Dirac operator $\dd_t$
gives rise to the minimal couplings to gravity and to
the non-Abelian gauge bosons and to the Yukawa
couplings to the Higgs boson that transforms like
$(2,-{\textstyle\frac{1}{2}},1)$. 

The spectral action $S_{CC}$ yields \cite{cc}, in
addition to the gravitational action, the entire bosonic
action of the standard model including the entire
Higgs sector with its spontaneous symmetry breaking.
The constraints for the coupling constants, $g_2^2=
g_3^2={\textstyle\frac{5}{3}}g_1^2
=3 \lambda$ occur because the Yang-Mills actions and
the $\lambda |\Phi|^4$ term stem from the same heat
kernel coefficient $f_4 a_4$. After renormalisation
through the big desert they yield a Higgs mass of 182
$\pm$ 17 GeV. 

\subsection{To gauge or not to gauge}

It is a long standing problem of the standard model
what symmetry of its fermion content do we gauge and
which one do we not gauge and there is no general
principle to answer this question. Not so in the
noncommutative setting where this choice is not
arbitrary. Indeed, the lifted automorphism group of
the internal part of the standard model is
\bb {\rm Aut}_{\hh_f}(\aa_f)=
SU(2)_L\times SU(3)_c\times U(N)_{qL}
\times U(N)_{\ell L}\times U(N)_{uR}
\times U(N)_{dR}\times U(N)_{eL},\ee
close to the identity. Only the isospin $SU(2)_L$, the
colour $SU(3)_c$ and two $U(1)$s in the five flavour
$U(N)$s are invited guests, i.e. they are images
under the lift $\ell$ of unitaries of the algebra
$\aa_f$. The subscripts indicate on which generation
multiplet the
$U(N)$s act, $qL$ for the $N=3$ left-handed quark
doublets, $\ell L$ for the left-handed lepton doublets,
$uR$ for the right-handed quarks of charge 2/3 and
so forth. The natural question at this point is: For
which of the 43 uninvited guests can we write letters
of invitation by extending the internal algebra? The
answer comes from the axioms of spectral triples, in
particular from the first order axiom and Poincar\'e
duality: only 10 additional symmetries can be gauged,
one left-handed and 9 right-handed ones:
\bb\aa_f=M_2(\cc)\op\cc\op M_3(\cc)\op M_N(\cc)
\owns (a,b,c,d)\ee
with three possible representations,
\bb \rho_{R}(b,d) := \pp{
d \ot 1_3&0&0\cr 0&\bar d \ot 1_3&0\cr 
0&0&\bar
b1_N},\ee\bb
\rho_{R}(b,d) := \pp{
b 1_N\ot 1_3&0&0\cr 0&\bar b 1_N\ot 1_3&0\cr 
0&0&\bar
d},\qq {\rm or}\qq
\rho_{R}(b,d) := \pp{
d \ot 1_3&0&0\cr 0&\bar d \ot 1_3&0\cr 
0&0&\bar
d},\ee
$\rho_L$, $\rho_L^c$ and $\rho_R^c$ being as in the
standard model. Only in the first of the three
possibilities, the $U(N)$ is anomaly free. A
phenomenological assessment of this
extension of the standard model with maximally
gauged flavour symmetry is under way.

\subsection{O'Raifeartaigh's reduction}

We owe to O'Raifeartaigh \cite{or} the intriguing
observation that all hypercharges in the standard
model conspire such that its group can be reduced to 
\bb G= SU(2)
\times U(1)\times
SU(3)/(\zz_2\times\zz_3),\ee where
\bb
\zz_2\times
\zz_3&=&\{(\exp[-k_22\pi i/2]1_2,\exp[k_22\pi
i/2+k_32\pi i/3],\exp[-k_32\pi
i/3]\,1_3),\cr &&\qq
 k_2=0,1,\ k_3=0,1,2\}\label{z2}\ee 
is the kernel of the representation of the standard
model on $\hh_L\op\hh_R$ given by equations
(\ref{hl}) and (\ref{hr}). The map 
\bb j:SU(2)\times U(1)\times SU(3)&\longrightarrow&
SU(2)\times U(3)\cr 
(s_2,u_1,s_3)&\longrightarrow&(s_2,u_1s_3)\\ 
 (s_2,(\det u_3)^{1/3}, (\det
u_3)^{-1/3}u_3)&\longleftarrow&
(s_2,u_3)\ee
defines an isomorphism from $G$ to $SU(2)\times
U(3)/\zz_2$.
 The latter form is useful
to reduce the group of unitaries
$U(\aa_f)=SU(2)\times U(1)\times U(3)$ to
$SU(2)\times U(3)$ by use of the unimodularity
condition. We write \cite{ls} this condition as an
injection
\bb m: SU(2)\times U(3)&\longrightarrow&
SU(2)\times U(1)\times U(3)=U(\aa_f)\cr 
(s_2,u_3)&\longmapsto&(s_2, \det u_3, u_3).\ee
Then the group representation of the standard model
 on $\hh_L\op\hh_R$ is $\ell\circ m\circ j$.
Imposing the unimodularity condition
 is equivalent to
imposing vanishing gauge and mixed
gravitational-gauge anomalies \cite{an}. Still today
the unimodularity condition remains a disturbing
feature of the noncommutative formulation of the
standard model but we must acknowledge that this
condition exists at all. It exists thanks to the
conspiration of the hypercharges that allows the
$\zz_3$ reduction. Now what about the $\zz_2?$
Remember the charge quantisation $q=\pm 2$ in
section 4. Its origin is clear. Although the algebra
representation $\rho$ is faithful by definition, the
group representation $\ell$ is not. Its kernel is
another $\zz_2\subset U(\aa)$. An immediate
calculation shows that  O'Raifeartaigh's
$\zz_2$ maps to this $\zz_2\subset U(\aa)$:
\bb &1_{\hh_L\op\hh_R}.&\cr 
&\uparrow\ \ell&\cr 
&(\exp[-k_22\pi i/2]1_2,\exp[k_26\pi
i/2],\exp[k_22\pi
i/2]\,1_3)&\!\!\!\!\!\!\!\!\!\!\!\!\!\!\!\!\!\!
=\exp[k_2i\pi](1_2,1,1_3)\cr 
&\uparrow\ m&\cr
&(\exp[-k_22\pi i/2]1_2,\exp[k_22\pi
i/2]\,1_3)&\cr 
&\uparrow\  j&\cr
&(\exp[-k_22\pi i/2]1_2,\exp[k_22\pi
i/2+k_32\pi i/3],\exp[-k_32\pi
i/3]\,1_3)&\ee 
Conversely, if the hypercharges had not conspired in
favour of O'Raifeartaigh's $\zz_2$ then the standard
model would not fit in Connes' geometrical frame.

\section{Dreams}

It is allowed to dream of a truly noncommutative
spectral triple with an algebra $\aa$ whose low
energy `approximation'
$E\ll\Lambda$ is the almost commutative
$\aa_t=\ccc^\infty (M)\ot(\hh\op\cc\op M_3(\cc))$.
`Truly noncommutative' means that all
automorphisms are inner,
Aut$(\aa)= $\,\,In$(\aa)= U(\aa)$. In this
situation the two lifts $L$ and $\ell$ coincide, see
figure, and the unification of gravity and Yang-Mills
forces would be perfect. The opposite extreme is  the
commutative algebra of pure gravity, $\ccc^\infty
(M)$, which has no inner automorphisms at all. The
lifted automorphisms of the truly noncommutative
triple would contain the spin cover of the Lorentz
group only approximately at low energies and we
could expect manifestations of the noncommutative
nature of spacetime in the form of violations of
Lorentz invariance above
$10^{17}$ GeV.  Amelino-Camelia has three
convincing arguments \cite{ac} that the experimental
observation of such violations might be possible
within the next ten years. The dream continues with
a generalization of the group of lifted automorphisms
of the truly noncommutative triple to a
Hopf algebra. And this Hopf algebra would be related to
a new quantum field theory which includes
gravity and which reduces to ordinary quantum field
theory at low energies. The mirage of this Hopf
algebra at low energies would be the one recently
discovered by Connes, Moscovici and Kreimer
\cite{cmk}.

\vskip 1truecm\noindent
As always, I am indebted to Raymond Stora. From him I
learnt the symmetric gauge some 17 years ago. It is
also a pleasure to acknowledge help and advice by
Samuel Friot, Bruno Iochum, Daniel Kastler, Serge
Lazzarini, Carlo Rovelli, Daniel Testard and Antony
Wassermann.

 \end{document}